\documentclass{llncs}
\usepackage{graphicx}
\usepackage{lscape}
\usepackage{amsmath}
\usepackage{subcaption}
\usepackage{mathtools}
\usepackage{float}
\usepackage{tabularx}
\usepackage{xcolor}
\usepackage{booktabs}

\newcolumntype{L}{>{$}l<{$}} 

\DeclarePairedDelimiterX{\infdivx}[2]{(}{)}{%
	#1\;\delimsize\|\;#2%
}

\title{As easy as 1, 2\ldots 4? \\
Uncertainty in counting tasks for medical imaging}

\begin{document}
	
\author{  Zach Eaton-Rosen\inst{1, 2} 
\and Thomas Varsavsky\inst{1, 2}	
\and Sebastien Ourselin\inst{2} 
\and M. Jorge Cardoso \inst{2} }

\institute{$^1$Centre for Medical Imaging Computing, University College London, London, UK\\
	$^2$Biomedical Engineering and Imaging Sciences, King's College London, UK\\
	}


\maketitle

\begin{abstract}
Counting is a fundamental task in biomedical imaging and count is an important biomarker in a number of conditions.
Estimating the uncertainty in the measurement is thus vital to making definite, informed conclusions. 
In this paper, we first compare a range of existing methods to perform counting in medical imaging and suggest ways of deriving predictive intervals from these. We then propose and test a method for calculating intervals as an output of a multi-task network. These predictive intervals are optimised to be as narrow as possible, while also enclosing a desired percentage of the data. We demonstrate the effectiveness of this technique on histopathological cell counting and white matter hyperintensity counting. 
Finally, we offer insight into other areas where this technique may apply. 



\end{abstract}

\section{Introduction}

Counting is a common analysis task required in a wide range of medical imaging applications from histology (cell counting) to neuroradiology (lesion counting). For any of these clinical biomarkers, accurate quantification of the degree of uncertainty over the measurements is of high importance in deciding an appropriate course of action. 
In this paper, we demonstrate an improved method for quantifying the uncertainty for counting tasks. 

Uncertainty can be broadly broken down into two constituent parts: model and data uncertainty. In the context of CNNs, `model' or `epistemic' uncertainty represents the uncertainty over the network (weights, hyperparameters, architecture) while `data' or `aleatoric' uncertainty represents the noise inherently associated with the data (noisy labels, measurement noise). 
Furthermore, out-of-distribution examples are likely to adversely affect the performance of machine-learning tools.

Epistemic uncertainty can be assessed through the comparison of several samples obtained from stochastic neural networks. If the stochasticity is induced by dropout, the sampling approximates full Bayesian inference~\cite{gal2016dropout}, which has been employed in image segmentation applications~\cite{kendall2015bayesian}. 
When training deep learning models, the stochasticity inherent in minibatching makes it possible to compare different models trained on the same training dataset: differing predictions of these models can be attributed to model uncertainty. 
A network can be trained to output diverse predictions from $m$ `heads'~\cite{lee2015m} coming from a common network trunk: the $m$ heads' differences are, again, due to model uncertainty. 

Heteroscedastic models of the noise uncertainty have been used in image super-resolution~\cite{tanno2017bayesian} and also exploited for spatially adaptive task loss weighting in multi-task learning~\cite{bragman2018uncertainty}. However, these parametric methods are restricted to unimodal, symmetric distributions, which are not necessarily realistic. Test-time augmentation has been used to perturb the data and thus infer the uncertainty from the differences in predictions~\cite{ayhan2018test,wang2019aleatoric}. In these approaches, the estimated uncertainty will depend wholly on the model's lack of invariance to the chosen augmentations: this may suggest that the models are undertrained or lacking capacity. 

While decomposing errors may be useful, other solutions exist. \textbf{Predictive Intervals} (PIs) estimate a lower and upper bound for an observation, such that the observation falls inside these bounds some chosen (high) percentage of the time: for a 95\% PI, we would expect 95 of 100 observations to lie within the interval.
PIs should satisfy the following properties: a) to be as small as possible, while b) still enclosing the appropriate fraction of results. This can be enforced through the loss function~\cite{pearce2018high}. In this work, we propose an extension of this method for application to counting tasks. 
We first describe different methods to perform counting tasks and assess uncertainty over measurements before presenting the loss function. We then describe the proposed amendment which is more flexible and stabler to train.

\section{Methods}
\subsection{Uncertainty in counting}

The overall aim of this work is to compute a predictive interval that a) minimises the interval width, whilst b) ensuring that the interval contains the appropriate percentage of results. Here we introduce several techniques to count cells and associated methods to calculate predictive intervals. While they present a novel contribution in their own right, these methods of counting are 
introduced here as a baseline against the method proposed in section \ref{sec:proposed}.
These baseline methods do not explicitly regress a predictive interval. Instead, we use the multiple outputs from the models (e.g. MC samples or $M$-heads) to sample predictions of count. Although we could simply use percentiles of these results to calculate intervals, these perform badly (the true count is often not within the obtained bounds). In order to mitigate this issue and introduce a fair comparison, the predictive intervals for each method below are calibrated post hoc following~\cite{eaton2018towards}: we transform the bounds affinely until they encompass a fraction $f > 1 - \alpha$ on the validation data. This transformation is then applied to the test-set estimates. 


\begin{figure}[ht]
	\centering
	\includegraphics[width=.9\linewidth, trim={0 0cm 0 0cm}, clip]{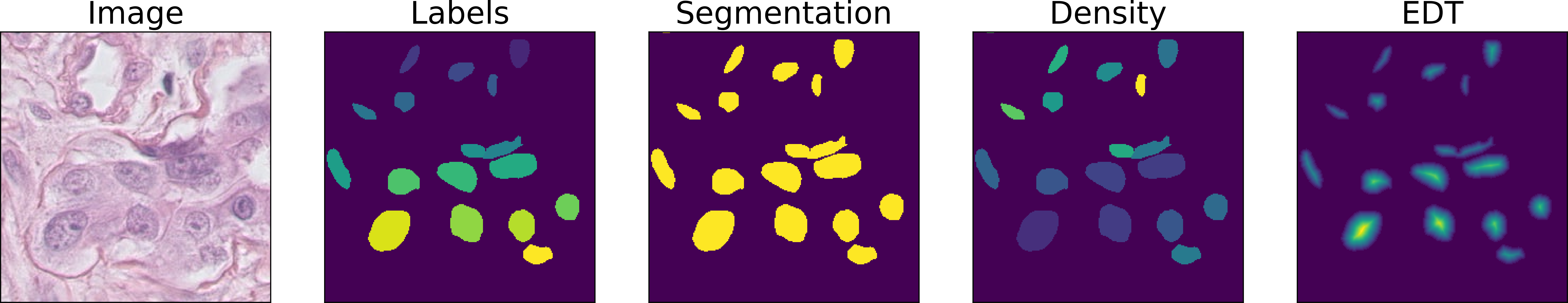}
	\caption{An illustrative figure of the cell data. The image has ground-truth labels that we binarise for the segmentation targets. The density and Euclidean Distance Transform (EDT) are targets for regression.}
	\label{fig:cell}
\end{figure}

\textbf{Segmentation-based:}
One counting method is to learn a segmentation of the input image and use connected-components analysis to determine the number of individual objects. 
To calculate uncertainty, we use three different approaches. We use Monte-Carlo samples of a network trained with dropout to produce N segmentation maps, counting the objects for each of the $N$. 
Secondly, we measure the number of objects at different thresholds of the output confidence map of the network. As the confidence threshold increases, fewer pixels will remain in the `foreground' class.  Finally, we also use $M$-heads~\cite{lee2015m}, which produces $M$ estimates of the segmentation with one forward pass of the network. This method produces a diverse mode-seeking ensemble, so higher variability in the heads may indicate that the model is more uncertain of the segmentation. For these methods, our target is the segmentation from Figure~\ref{fig:cell}. 

\textbf{Regressing the Euclidean Distance Transform:}
Naylor et al~\cite{naylor2019segmentation} introduced a regression-based method for cell counting. Given an input image, the network learns to approximate the Euclidean Distance Transform of the cell segmentation maps (see Figure~\ref{fig:cell}). A non-maximum suppression is then used to count the cells. In order to calculate the uncertainty of the cell count, we use the MC-sampling paradigm described above, with a network with dropout trained.

\textbf{Regressing the pixel-wise cell density:}
One popular technique for counting in computer-vision applications is to use a regression formulation to estimate a density-map from the raw image: summing over all pixels returns the count~\cite{lempitsky2010learning,xie2018microscopy}. 
The density estimation function we use is a convolutional neural network. For these experiments, the ground truth density map has value $0$ for background, and $\frac{1}{n_i}$ otherwise, where $n_i$ is the number of pixels in the $i^{th}$ object. 
The network we use to estimate the density is a multi-task network with a shared backbone and two `heads'. One head learns the segmentation while the other learns the density map. 
For uncertainty estimation, we use the $M$-heads paradigm to introduce variance in the output.

\subsection{Distribution-free uncertainty estimation}
The previous examples all rely on sampling, followed by post-training calibration step which maps the sample uncertainty to the target predictive bounds. Conversely, the authors in~\cite{pearce2018high} proposed to regress the predictive intervals directly from the data. They aim to estimate lower and upper confidence bounds for a desired quantity $y$, where $b_l$ and $b_u$ are the lower and upper bounds respectively. The hyper-parameter $\alpha$ determines the desired width of the interval: it is defined such that:

i) $ p\big(y \in [b_l, b_u]\big) = 1-\alpha$, and 

ii) $p\big(y \in \cup \big\{ (b_u, \infty], [-\infty, b_l)\big\} \big) = \alpha$.


Common choices for $\alpha$ include 0.10, 0.05 and 0.01, representing $90\%$, $95\%$ and $99\%$ confidence intervals respectively. 
In the original work, the authors propose a loss function $\mathcal{L}_{QD}$ to estimate `Quality-Driven', distribution-free predictive intervals. For any given datapoint, $x_i$, the model returns $b_{l_i}$ and $b_{u_i}$. For each input $x_i$ we assess whether the observed corresponding datapoint $y_i$ is in or out of the prediction bounds $[b_{l_i},b_{u_i}]$. In order to provide useful information on the behaviour of the predictive interval estimates over multiple examples, the loss function is allowed to reason over an entire minibatch of size $n$. In this setup, an indicator variable is used to express if $y_i$ within the predictive interval or not. The number of times $y_i$ falls within the predictive interval is given by a binomial distribution 
$\operatorname{Binomial}(n,(1-\alpha))$, assuming i.i.d. data, which can be approximated by a normal distribution for large $n$. 
The loss,  $\mathcal{L}_{QD}$, is then defined as the sum of a width term and the log-likelihood term. 

\begin{equation}
\label{eqn:qd_loss}
\mathcal{L}_{QD} = \bar W_{captured}  + \lambda \frac{n}{\alpha(1-\alpha)} \text{max}(0, (1 - \alpha) - q) ^ 2
\end{equation}

\noindent where $\bar W_{captured}$ is the mean interval width for intervals that capture their associated ground truth, $\lambda$ is a constant, $n$ is the number in a batch, and $q$ is the fraction of points that lie within their estimated predictive interval bounds. 

\subsection{Proposed Extension}
\label{sec:proposed}

In practice, we found $\mathcal{L}_{QD}$ difficult to optimise. We observed periodic instabilities in the training and attributed this, in part, to the one-sided nature of the second term; we sought to modify its formulation appropriately. 
With this in mind, let $P_s$ be a discrete probability distribution function representing the probability of being `in' or `out' of the predicted interval, where the subscript $s$ denotes `state', i.e. $P_{in} = 1-\alpha$,  $P_{under} = \alpha/2$ and $P_{over} = \alpha/2$. 
For any minibatch, we define the observed proportions of `over', `under' and `out' samples as $Q_s$, and use the Kullback-Leibler divergence (KL) to enforce similarities between the target $P$ and the observed $Q$.  With this framework, $Q$ could be encouraged to match any desired distribution $P$ (for instance, estimating several bounds to correspond to percentiles). Note that, in this proposed framework, and contrary to \cite{pearce2018high}, $P$ can represent any chosen distribution.

\begin{equation}
\label{eqn:loss}
\mathcal{L}_{distribution} = KL(P||Q) = \sum_s{P_s log\frac{P_s}{Q_s} } = \sum_s \left \{{P_slog(P_s) - P_slog(Q_s)} \right\}\\
\end{equation}

Since the target distribution $P$ is constant, minimizing $L_{distribution}$ is identical to minimising the cross-entropy term with respect to the network weights; this means that we are simply promoting that the proportion of inliers in a given minibatch matches our desired distribution.

As this loss function uses the categorical membership of the $y_i$ to estimate $Q$, 
a soft membership function is used to make this operation differentiable and hence suitable for back-propagation.
We calculate the proportion inside the bounds as 
$
Q_{in,soft}=
\sigma\left(\xi\left(\mathbf{y}-{\mathbf{b}}_{\mathbf{l}}\right)\right) \odot \sigma\left(\xi\left({\mathbf{b}}_{\mathbf{u}}-\mathbf{y}\right)\right)
$.
The minibatch of ground truth counts $\mathbf{y}$ is compared with the regressed bounds, $\mathbf{b}_{\mathbf{l}}$ and $\mathbf{b}_{\mathbf{u}}$, with $\sigma$ representing the sigmoid function and $\xi$ being a positive softening constant (set to $\xi=2$). This formulation of the soft boundaries is as in~\cite{pearce2018high}, with the other soft memberships (over, under) being set analogously.

Our proposed loss is given by: 
\begin{equation}
\mathcal{L}_{proposed} =  \bar W_{captured}  + \lambda \sum_s{P_slog(Q_s)}
\end{equation}
\noindent In short, instead of using a one-sided data likelihood term, we used the cross-entropy calculated between the chosen $P$ and $Q$.  This reformulation has not only added flexibility, allowing for different state chosen $P$ distributions, but we also found it easier to train than the model in Eq.~\ref{eqn:qd_loss}.

\subsection{Network Architectures and Implementation Details}

The U-Net~\cite{ronneberger2015u} forms the basis of all of our CNNs. The multi-task network has a U-Net backbone with the same parameters as for the single-task approaches. It then splits into separate branches (one for segmentation and one for density prediction). 

\begin{figure}[htb]
	\centering
	\includegraphics[width=1\linewidth, trim={0 0cm 0 0cm}, clip]{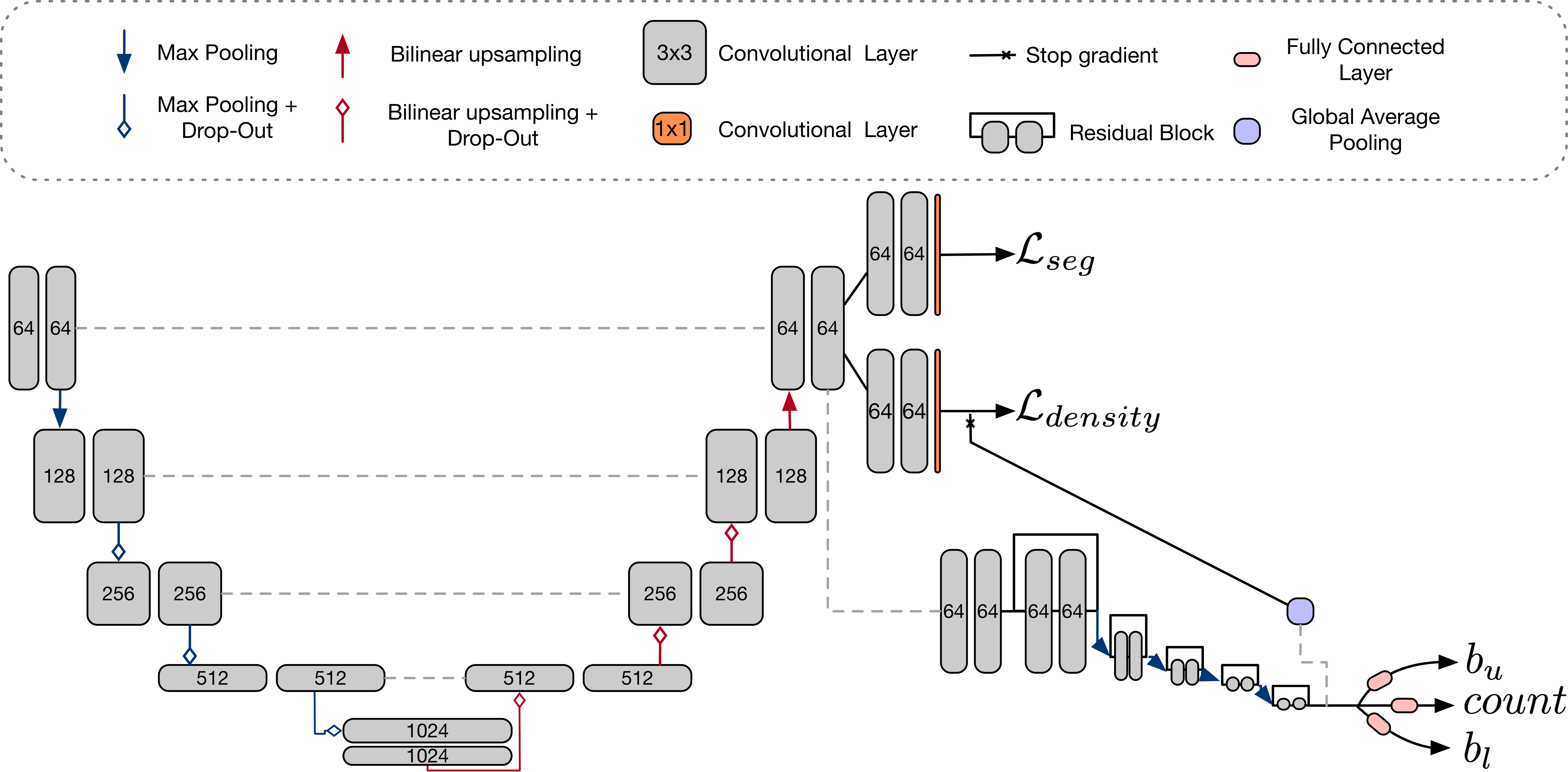} 
	\caption{Multi-task architecture for simultaneous segmentation and uncertainty prediction. All convolutions are 3$\times$3 by the channel width, denoted in the diagram. The U-Net is complemented by an `arm' which has residual blocks followed by max-pooling (maintaining 64 filters) until it reaches the output layer, where it returns an upper and lower bound. Dropout is enabled for methods that require MC-sampling where indicated in the diagram, with $p=0.5$. The bounds are trained with the loss from Equation~\ref{eqn:loss}.  
	}
	\label{fig:unet}
\end{figure}

For the proposed method of uncertainty prediction, we fit a regression network with three output quantities. First, it outputs a `predicted count' which is trained with an  $\mathcal{L}_2$ loss. The other two quantities are the upper and lower bounds, for a given $\alpha$. In our experiments we choose $\alpha=0.1$, making them 90\% intervals. The architecture was chosen to have residual blocks as part of the arm to avoid vanishing gradients. 

Due to the complexity of our network, we train it in stages. First, we train the U-Net part on the segmentation and density tasks. We then freeze the weights and train the predictive intervals, with a batch size of 64: as discussed, a large batch size to estimate good batch-wise statistics. 
In our experiments, we set $\lambda = 30$. The auxiliary $L_2$ loss is set to 1e-3. We parametrise the outputs of the network as such: the estimate for the mean value has no final activation. The upper and lower residuals go through a softplus activation function and are then added or subtracted, as appropriate, to the mean estimate. The segmentation has a sigmoid output, and the density a square function. Models are trained with early stopping as determined on the validation set, in NiftyNet~\cite{gibson2018niftynet}. 
 
\subsection{Data}

The proposed counting methods are applied to the counting of cells from histological slides~\cite{naylor2019segmentation}. This dataset has 33 labelled slides of dimension $512\times512$, taken from 7 different types of tissue, and each slide has an associated cell label map used here as the count ground truth.
We separated this into 7 `test' images (21\%), one from each cell type, and 4 `validation' images. 
To have larger batch sizes, we trained on images of size $256\times256$ and hence quartered each image while keeping the same fold label. Heavy augmentation is applied to the images in the training set (see figures in Supplementary Materials) using the `imgaug' library~\cite{imgaug}. 

We also fit to a white-matter hyperintensity (WMH) dataset~\cite{kuijf2019standardized}. In this task, we demonstrate a slightly different parameterisation of our bound prediction. We fitted an $M$-Heads model to the WMH segmentation and used the same model as a feature extractor to train the predictive bounds. In this data, of the 60 subjects, we used an 80/10/10 split for training/testing/validation respectively.

\section{Results}
\label{sec:results}

\setlength{\tabcolsep}{10pt}

\begin{table}[htbp]
	\resizebox{\columnwidth}{!}{
\begin{tabular}{@{}llcccccccc@{}}
	\toprule
	Method & Paradigm & \multicolumn{1}{l}{C$_{EST}$} & \multicolumn{1}{l}{MAE $\pm$ STD} & \multicolumn{1}{l}{$\rho$} & \multicolumn{1}{l}{$f_{uncal}$} & \multicolumn{1}{l}{$W_{uncal}$} & \multicolumn{1}{l}{$f_{cal}$} & \multicolumn{1}{l}{$W_{cal}$} \\ \midrule
	Segmentation & Thresholds  & 30.32 & 6.16 $\pm$ 5.43 & 0.90 & 0.61 & 15.33 & 0.86 & 23.42 \\
	& M-Heads &  25.14 & 2.83 $\pm$ 3.34 & 0.95 & 0.46 & 2.80 & 0.96 & 102.22 \\
	& MC samples & 31.36 & 6.70 $\pm$ 6.37 & 0.87 & 0.43 & 7.25 & 0.89 & 27.82 \\
	\begin{tabular}[c]{@{}l@{}}EDT \\ Regression:\end{tabular} & MC samples & 26.16 & 2.85 $\pm$ 2.03 & 0.97 & 0.25 & 3.00 & 0.96 & 48.09 \\
	& \% Errors & --- & --- & --- & --- & --- & 0.82 & 18.61 \\
	\begin{tabular}[c]{@{}l@{}}Density \\ Regression\end{tabular} & M-Heads & 26.42 & 3.53 $\pm$ 3.12 & 0.96 & 0.57 & 6.42 & 0.75 & 15.64 \\
	$\mathcal{L}_{QD} $ & PI-estimate & 26.01 & 3.04 $\pm$ 2.82 & 0.96 & --- & --- & 1.0 & 29.31 \\
	\textbf{Ours} & PI-estimate & 26.23 & 2.93 $\pm$ 2.93 & 0.96 & --- & --- & 0.93 & \textbf{12.20} \\ \midrule
	\begin{tabular}[c]{@{}l@{}}Lesions: \\ Segmentation\end{tabular} & M-Heads & 6.08 & 2.89 $\pm$ 2.96 & 0.83 & 0.09 & 0.63 & 0.96 & 24.1 \\
	\textbf{Ours} & PI-estimate &6.08 & 2.89 $\pm$ 2.96 & 0.83 & {---} & {---} & 0.89 & \textbf{10.93} \\ \bottomrule
\end{tabular} 
}
	\caption{ 
		C$_{EST}$ is the average estimated count. The ground truth counts were 25.71 for cells and 8.19 for lesions. MAE is the mean absolute error, $\pm$ standard deviation. $\rho$ is the correlation coefficient between estimated and ground truth counts. 
		$f$ is the fraction of the ground truth points within the bounds and $W$ is the mean width of the intervals --- evaluated on both uncalibrated and calibrated intervals. }
	\label{tab:res}
\end{table}

\begin{figure}[bp]
	\centering
	\includegraphics[width=1.\linewidth]{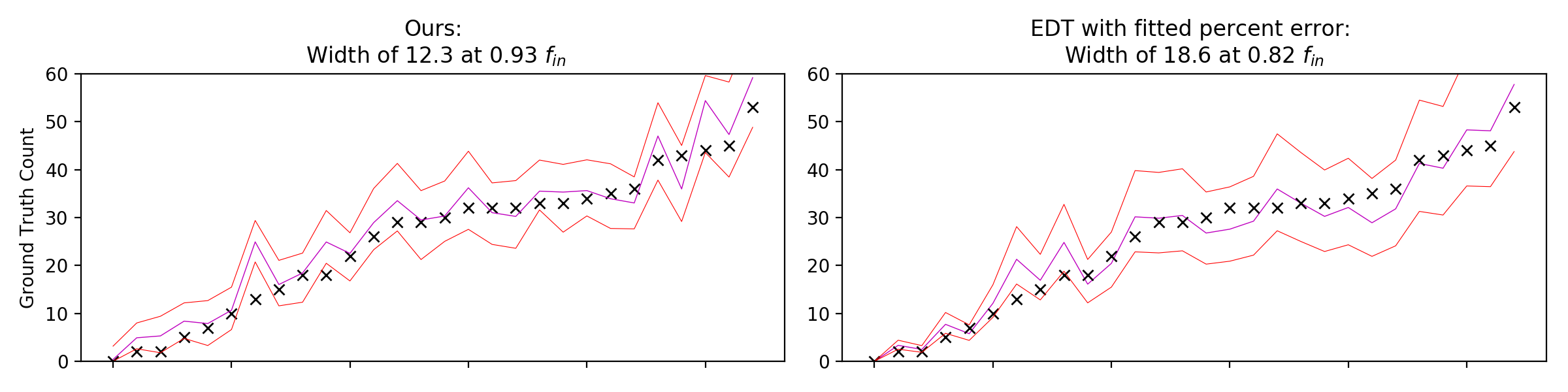}
	\caption{Here we contrast our model (left) with the model with fitted percentage noise. The points represent the ground truths, and the lines represent the upper bound, mean and lower bound (note that the lines are for ease of visualisation: the $x$-axis is not continuous).}
	\label{fig:my_ints}
\end{figure}

We show the results in Table~\ref{tab:res}. 
All of the models exhibit good performance in counting, with the correlation between predicted and GT counts being above 0.8 for all models. 
The uncalibrated predictive intervals capture anything from 9\% to 61\% of the data. After calibrating these models, many achieved the correct percentage of inliers for the predictive intervals. Some (for example, the $M$-heads density regression with $f_{in}=0.75$) did not: this may indicate that the calibration methods were overfitting (despite having few parameters --- only an affine transformation). 
Our model predicted significantly smaller interval widths than for the baseline methods for both cells and lesions. 
While the baseline methods may seem to give large bounds, in the cases of cells, there may be a count of over 100 per image and in the lesions, up to 35. Because the EDT regression had the lowest MAE, we chose another, simple, baseline: we simply had a percentage count as the error. This method achieved a width of 18.6, compared to 12.20 (ours) in Figure~\ref{fig:my_ints}. It also does not capture the desired percentage of inliers, as it is too small.
For the model fitted with $\mathcal{L}_{QD}$, we report the best results obtained after 3 independent model-fits, as we found the loss was unstable to fit --- however, it still underperformed our proposed loss function.

\section{Discussion}
\label{sec:disc}

The aim of this work was to accurately predict intervals, such that they were of minimal width while containing the desired numbers of ground truth values. Our predictive bounds were over 20\% smaller than the nearest competitor method while retaining the correct number of inliers; these smaller bounds are correspondingly more informative. We have demonstrated these results on a cell histology dataset and on WM lesions.
For the cells, the next-best method was applying a constant percentage uncertainty to the counts of the EDT regression framework. Fitting this percentage is, in essence, optimising the same loss as we applied, but only using the predicted counts (and none of the image features). 
The fact that our model outperforms this baseline implies that the imaging features are being used to make an informed estimate of predictive error. 

One limitation of this work is that it is not likely to generalise to samples drawn from outside of the training distribution. Domain-adaptation methods could help ameliorate this. 
It is also not an interpretable model and hence it would of interest to use model introspection methods to investigate how the network decides on its bounds.
As the model we have presented can, in principle, be applied to any estimate derived from a machine-learning model, future work will investigate its applicability to 3D counting problems and a wider range of clinical biomarkers.

%
%
%



%

\textbf{Acknowledgements:} ZER is supported by the EPSRC Doctoral Prize. MJC \& SO are supported by the Wellcome Flagship Programme (WT213038/Z/18/Z) and the Wellcome EPSRC CME (WT203148/Z/16/Z). 
We gratefully acknowledge NVIDIA Corporation for the donation of hardware.

\bibliography{./bib}

\appendix

\section{Supplementary Materials}\label{supp}
\subsection{Data Augmentation}\label{supp:aug}

\begin{figure}[h]
	\centering
	\includegraphics[width=1.\linewidth]{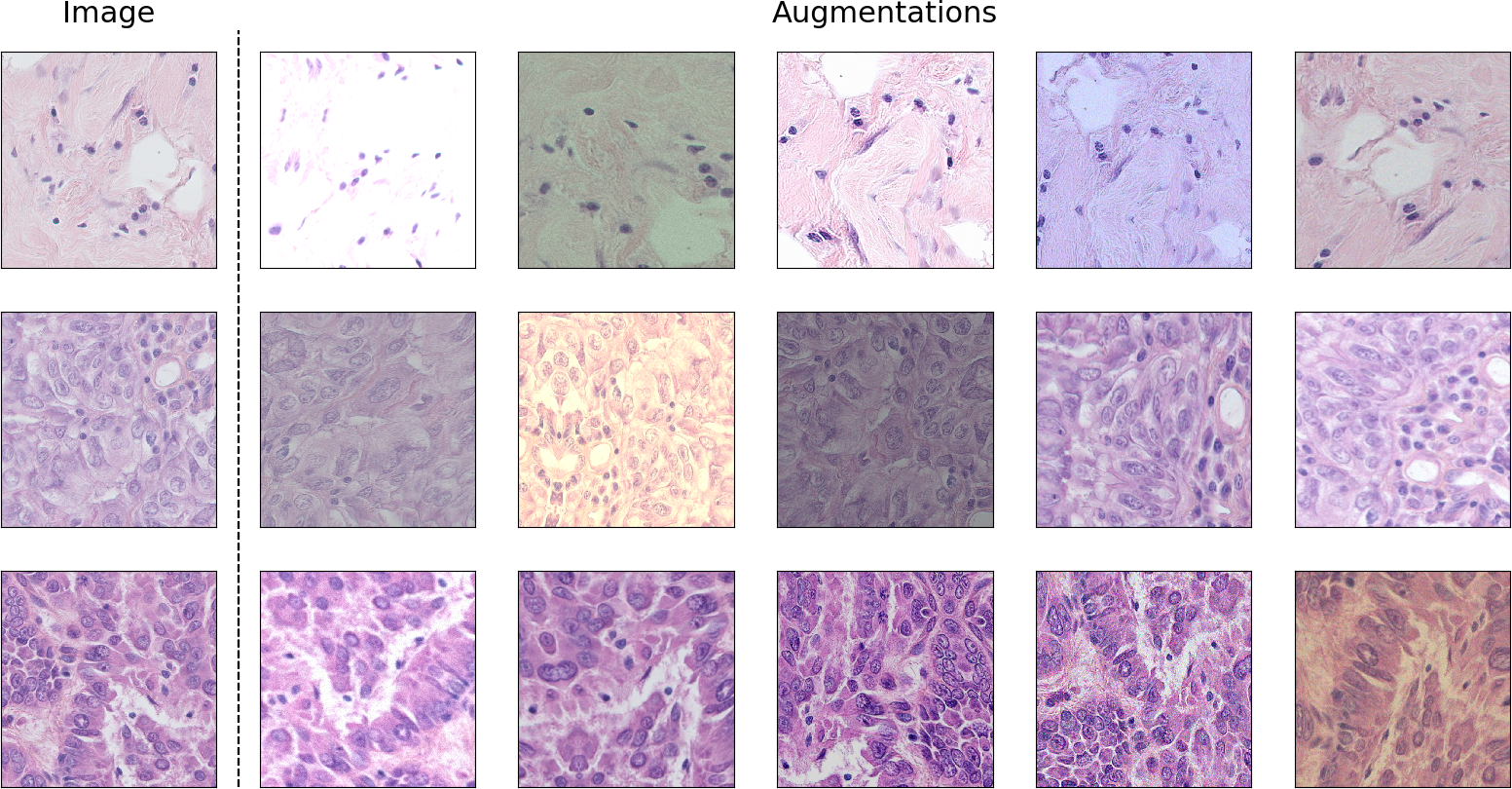}
	\caption{Examples of augmented cell data}
	\label{fig:aug_cells}
\end{figure}

\begin{figure}[]
	\centering
	\begin{tabular}{@{}llr@{}}
		\toprule
		Augmentation            & Details                                                                                     & \% applied to\\ \midrule
		Flip                    & Left-Right. Up-Down                                                                         & 50, 50             \\
		Random Cropping         & \begin{tabular}[c]{@{}l@{}}Crops $\in (0, 0.1)$ of \\ image dimension.\end{tabular}         & 100                \\
		Gaussian Blur           & $0 < \sigma < 2$, chosen at random                                                          & 50                 \\
		Piecewise Affine        & Scale $\in (0.02, 0.07)$                                                                    & 50                 \\
		Contrast Normalisation  & Contrast $\in (50, 150\%)$                                                                  & 100                \\
		Sharpening              & \begin{tabular}[c]{@{}l@{}}alpha $\in (0, 0.6)$, \\ lightness $\in(0.75, 1.25)$\end{tabular}& 50                 \\
		Random Additive Noise   & Per pixel noise $\in (-30, 30)$                                                             & 100                \\
		Gaussian Additive Noise & Scale of $5\%$                                                                              & 100                \\
		Random Brightening      & $0.8 <$ brightness $< 1.2$                                                                  & 100                \\
	    Random Scaling          & $0.8 < $ scale $ < 1.2$                                                                         & 100                \\
		Random Rotation         & $-35 < \Theta < 35$                                                                         & 100                \\
		Random Translation      & $x_{trans}, y_{trans} \in (-.15, .15)$                                                      & 100                \\
		Shearing                & (-8, 8)                                                                                     & 100                \\ \bottomrule
	\end{tabular}
\caption{The augmentation for the cell images was done using the `imgaug' GitHub repository~\cite{imgaug}. All augmentation was performed offline for computational efficiency.  This table shows our chosen settings.}
\label{tab:aug}
\end{figure}

\end{document}